\documentclass[12pt]{article}
\usepackage{hyperref}
\usepackage{amsmath}
\usepackage{graphicx}
\usepackage{amssymb} 
\newcommand{\Mat}[1]{{{\boldsymbol{#1}}}}
\newcommand{\abs}[1]{\left\vert#1\right\vert}
\def\be{\begin{equation}}
\def\ee{\end{equation}}
\def\bea{\begin{eqnarray}}
\def\eea{\end{eqnarray}}
\def\dd{\mathrm{d}}
\title{Main effects of the Earth's rotation on the stationary states of ultra-cold neutrons}

\bigskip

\author{
Mayeul Arminjon\\
\small\it Laboratoire ``Sols, Solides, Structures, Risques'' (CNRS \& Universit\'es de Grenoble),\\
\small\it BP 53, F-38041 Grenoble cedex 9, France.\\
} 
\date{ }
\begin{document}

\maketitle


\begin{abstract}
\noindent 
The relativistic corrections in the Hamiltonian for a particle in a uniformly rotating frame are discussed. They are shown to be negligible in the case of ultra-cold neutrons (UCN) in the Earth's gravity. The effect, on the energy levels of UCN, of the main term due to the Earth's rotation, i.e. the angular-momentum term, is calculated. The energy shift is found proportional to the energy level itself.\\

03.65.Pm 	Relativistic wave equations\\ 
04.25.Nx 	Post-Newtonian approximation; perturbation theory; related approximations\\

\end{abstract}

\section {Introduction}
The Granit experiment has verified the quantization of the energy levels of ultra-cold neutrons (UCN) in the Earth's gravity field, in measuring the transmission of UCN through a horizontal slit \cite{Nesvizhevsky2003}. New, more precise experiments are being foreseen, in order to accurately determine the energy levels \cite{Nesvizhevsky2005}. The aim of this paper is to calculate the main effects of the Earth's rotation on the gravitational energy levels of UCN. The effect of the Earth's rotation on the quantum-mechanical phase of neutrons has been detected in the neutron interferometry experiments of Werner {\it et al.} \cite{WernerStaudenmannColella1979}. Their theoretical interpretation is based on the non-relativistic Hamiltonian for a neutral particle in the inertial/gravitational field of the uniformly-rotating Earth:
\be\label{H-op}
\mathrm{H} = \frac{{\bf \hat {p}}^2}{2m}- \Mat{\omega}{\bf.L}+V_0,\qquad {\bf L \equiv r \wedge \hat {p}},\qquad {\bf \hat {p}}\equiv -i\hbar \nabla, 
\ee
with $V_0\equiv -mU_\mathrm{N}$, where $U_\mathrm{N}({\bf r})$ is the positive Newtonian potential, and where $\Mat{\omega}$ is the constant rotation velocity vector. The gravity acceleration is ${\bf g}=+\nabla U_\mathrm{N}$, thus $V_0\simeq -m{\bf g.x}$  in the locally-uniform Earth gravity, where ${\bf x \equiv r-r}_0 $, with ${\bf r}_0$ taken, say, on the lower side of the slit and at its entrance, in the Granit setting (which is described in Refs. \cite{Nesvizhevsky2003,WestphalAbele2006,VoroninAbele2006}).\\

In Sect. \ref{DiracHamiltonian}, we shall take into account the relativistic effects which occur for a Dirac particle, in order to see if they are large enough to be detected. To this aim, we shall use mainly a study, by Hehl \& Ni \cite{HehlNi1990}, of the post-Newtonian Hamiltonian of a Dirac particle in a purely inertial field, and also a previous work \cite{A38} on the energy levels of a Dirac particle in a purely gravitational field. In the last part (Sect. \ref{UCN}), we shall apply these results to the stationary states of UCN in the Granit experiment: we shall first estimate the relativistic effects, and then we shall calculate the main effect of the Earth's rotation, which is the (non-relativistic) effect of the angular-momentum term.\\

\section{Hamiltonian for a Dirac particle in an Earth-based laboratory} \label{DiracHamiltonian}
\subsection{Statement of the problem}
The quantum equation for an elementary relativistic particle with spin $\frac{1}{2}$ in a flat spacetime is the Dirac equation. In the case with a gravitational field, which is described by a curved spacetime, the Dirac equation is usually modified to the form derived independently by Weyl and by Fock in 1929, hereafter the Dirac-Fock-Weyl (DFW) equation. Being a generally-covariant equation, the DFW equation may also be used to describe the effects of a non-inertial frame. However, for a neutral particle obeying Dirac's equation, the presence of an electromagnetic field has no effect whatsoever. This is contrary to the observed existence of a magnetic moment for the neutron, detected in magnetic scattering experiments: these gave the first indication that the neutron might be non-elementary, as is of course admitted in the framework of the quark model.\\

Nevertheless, we may admit as a working assumption that, in the absence of an electromagnetic field, the inertial/gravitational effects on neutrons are correctly described by the DFW equation. Our task then is to look for an approximate Hamiltonian, suitable to describe accurately enough the effects of a weak inertial/gravitational field on a slow particle. There has been some amount of work devoted to the subject (see {\it e.g.} \cite{deOliveiraTiomno1962,VarjuRyder1998,Obukhov2001,SilenkoTeryaev2005}), even though, as noted by Varj\'u and Ryder \cite{VarjuRyder2000}, ``there is unfortunately some disagreement among [the different] papers." This may be partly explained by noting that the operator transformations of the Foldy-Wouthuysen type, which are used in Dirac theory to select the positive-energy states, are not uniquely defined \cite{SilenkoTeryaev2005}. In addition: {\bf i}) the approximation scheme was not made explicit in these works, and {\bf ii}) these works were concerned with the Hamiltonian for the four-dimensional complex wave function, whereas it turns out that, as for the flat-spacetime Dirac equation, the solutions of the stationary DFW equation have only two independent complex components \cite{A38}. In the asymptotic post-Newtonian scheme used in Ref. \cite{A38}, the positive-energy states are automatically selected by assuming that any solution $E^{(\lambda )}$ of the eigenvalue problem for the stationary energy levels should have an expansion 
\be \label{E-mc2} 
E^{(\lambda )}= mc^2 + O(\lambda ),
\ee
where $\lambda \ll 1$ is the gravitational field-strength. In that way, one avoids the ambiguity which may occur in the Foldy-Wouthuysen-type transformations. This study leads to an explicit equation for the stationary energy levels in a gravitational field, which shows that, for UCN, the corrections brought by the DFW equation to the Schr\"odinger equation in the potential $mgz$ are utterly negligible \cite{A38}. This result has been since confirmed by an independent study \cite{Boulanger-Spindel2006}.\\

However, these two studies \cite{A38,Boulanger-Spindel2006} were based on the assumption of a static metric (as were the works \cite{VarjuRyder1998,Obukhov2001,SilenkoTeryaev2005}). Hence, they do not involve the effect of the Earth's rotation, because the transition to a rotating frame implies a non-diagonal metric, $g_{0j} \ne 0$. One might try to adapt the foregoing work \cite{A38} to a such ``stationary metric."  Our aim here is to estimate the ``relativistic" corrections, {\it i.e.,} the difference between the predictions got either with the DFW equation or with the non-relativistic Schr\"odinger equation based on the Hamiltonian (\ref{H-op}). Anticipating the smallness of these corrections, we may consider separately {\bf i}) the ``purely gravitational" corrections, i.e., those corrections which would be there if the Earth's gravitational field was static---the foregoing studies \cite{A38,Boulanger-Spindel2006} showed that these corrections are negligible for UCN---and {\bf ii}) the ``purely inertial corrections", which would be there if the rotating Earth had no gravitational field. The latter ones can be obtained from the work by Hehl and Ni \cite{HehlNi1990}, which is devoted to the inertial effects in a flat spacetime.

\subsection{Post-Newtonian Hamiltonian in a non-inertial frame in flat spacetime}

The work \cite{HehlNi1990} is based on considering a (point-like) observer subjected to non-inertial (accelerated) motion in {\it flat} spacetime, thus in the absence of a gravitational field; due to the assumed rotation of an orthonormal triad $({\bf e}_i)$ that the observer carries with him, this approach also takes into account the rotation of the local reference frame. According to Kiefer and Weber \cite{KieferWeber2005}, this approach is able to describe also the {\it local} effect of the gravitational field, by substituting $-{\bf g}$ for the three-acceleration ${\bf a}$.
\footnote{\
These authors write: ``by replacing ${\bf a}$ with ${\bf g}$." Of course, the correct sign depends on which acceleration one refers to. Recall that, in a non-inertial frame F in non-relativistic mechanics, the inertial force on a particle is ${\bf F}_\mathrm{i}=m({\bf a}-{\bf a}_0)$, where ${\bf a}$ is the acceleration of the particle in F and ${\bf a}_0$ is its acceleration in an inertial frame---allowing to write Newton's second law in F as ${\bf F}\equiv {\bf F}_\mathrm{i}+{\bf F}_0=m{\bf a}$, with ${\bf F}_0$ the force seen in an inertial frame. A particle that remains at rest in F (relative velocity ${\bf u}={\bf 0}$, and ${\bf a}={\bf 0}$) is hence subjected to the inertial force ${\bf F}_\mathrm{i}=-m{\bf a}_0({\bf r},{\bf u}={\bf 0})$. Thus, the absolute acceleration field ${\bf a}_0({\bf r},{\bf u}={\bf 0})$ of the observers bound to F plays the role of the {\it opposite} of a gravitational attraction field, ${\bf g}=-{\bf a}_0$. 
}
 Hehl \& Ni \cite{HehlNi1990} define several coordinate systems and tetrads, one of which, $(x^\mu)$ [following the rotation of the triad $({\bf e}_i)$ and the corresponding tetrad $({\bf e}_\mu )$] being qualified thus: ``Such a local coordinate system is what we actually use in our laboratory." Writing the DFW equation in that system, and using three successive Foldy-Wouthuysen transformations, they get the following form of the DFW equation \cite{HehlNi1990} (writing $\beta\equiv\gamma^0$, the standard-representation Dirac matrix):
\be \label{Schroed-HFW}
i\hbar\frac{\partial\psi}{\partial t}=\mathrm{H}_{\rm FW}\psi\ ,
\ee
with
\bea \label{HFW}
\mathrm{H}_{\rm FW} &=& \beta mc^2 -\frac{\beta\hbar^2}{2m}\Delta 
+ \beta m({\mathbf a.} {\mathbf r})-\Mat{\omega}.{\mathbf L}-\Mat{\omega}.{\mathbf S}-\frac{\beta\hbar^2}{2m}{\nabla}\frac{\mathbf{a.r}}{c^2}.{\nabla}
  \nonumber\\
& & 
-\frac{i\hbar^2}{4mc^2}\Mat{\sigma }.({\mathbf a}\wedge 
{\nabla})+ \mathrm{higher-order\ terms},
\eea
with ${\mathbf S} \equiv \frac{1}{2}\hbar\Mat{\sigma}$, and where $\Mat{\sigma}$ denotes the space ``vector" made with the Pauli matrices $\sigma ^j$---or rather, with the matrices 
\\
\be
\Sigma ^j \equiv \left (\begin{array}{cc} \sigma ^j \ 0 \\ 0\ \sigma ^j \end{array} \right),
\ee
since here the wave function $\psi $ has four complex components. In Ref. \cite{KieferWeber2005}, an additional term $-\frac{\beta\hbar^4}{8m^3c^2}\Delta^2$ is present, and the remaining higher-order terms are stated to be $O(c^{-3})$. Moreover, in the last term above, $\Mat{\sigma }.({\mathbf a}\wedge 
{\nabla})$ is replaced by $\beta\Mat{\sigma }.({\mathbf a}\wedge 
{\nabla})$, which has the same effect on the ``positive-energy spinor" $\tilde{\varphi}$, in a decomposition $\psi =(\tilde{\varphi} ,\tilde{\chi} )$ of the four-components spinor $\psi $. Indeed, we have
\be
\beta \psi = (\tilde{\varphi} ,-\tilde{\chi} ).
\ee
Thus, the second, third and fourth terms in the post-Newtonian DFW Hamiltonian (\ref{HFW}) correspond exactly to the non-relativistic Hamiltonian (\ref{H-op}), if we put ${\bf a=-g}$. This must be the case, since the other terms in (\ref{HFW}) are either the spin effect $-\Mat{\omega}.{\mathbf S}$, necessarily absent with the scalar Hamiltonian (\ref{H-op}), or the relativistic effects. However, it is not necessary for us to mimic the gravitational field in a flat spacetime, since in this subsection we want to describe specifically the relativistic corrections to the non-inertial effects in a flat spacetime. Thus, what we actually should recover in the non-relativistic limit is the non-relativistic Hamiltonian (\ref{H-op}) \underline {with $V_0=0$}, corresponding to \underline {${\bf a=0}$ in (\ref{HFW})}. It means that, for the application of the work \cite{HehlNi1990}, the ``non-inertial observer" is on the axis of rotation of the Earth, and his ``local triad" is just the one that rotates as does the Earth. 

\section{Application to the stationary states of UCN in the Earth's gravity}\label{UCN}
\subsection{Estimate of the relativistic effects}

Searching for stationary solutions $\psi (t,{\bf r}) = \phi (t)\,A({\bf r})$ of the Schr\"odinger-like equation (\ref{Schroed-HFW}), one gets energy states:
\be\label{energy-state}
 \mathrm{H}_{\rm FW}A=E\,A.
\ee
Setting
\be\label{phi-chi}
A({\bf r})=(\varphi ({\bf r}),\chi ({\bf r})),
\ee
we have explicitly from (\ref{HFW}), for the independent 2-spinor $\varphi $:
\be \label{2-spinor-HFW}
\mathrm{H}_{\rm FW}\varphi=\left[mc^2-\frac{\hbar^2}{2m}\Delta 
+  m({\mathbf a.} {\mathbf r})-\Mat{\omega}.({\mathbf L}+{\mathbf S})-\frac{\hbar^2}{2mc^2}\left(\mathbf{a.r}\Delta +{\bf a.}\nabla +\frac{i}{2}\Mat{\sigma }.({\mathbf a}\wedge {\nabla})\right)\right]\varphi.
\ee
Let us briefly compare Eq. (\ref{2-spinor-HFW}), based on the work \cite{HehlNi1990}, to other recent approaches to the positive-energy PN Hamiltonian for a Dirac particle. Thus, we may compare (\ref{2-spinor-HFW}) with Eq. (52) of Boulanger {\it et al.} \cite{Boulanger-Spindel2006}, also based on a Foldy-Wouthuysen transformation, by assuming $\Mat{\omega}={\bf 0}$ here (since the rotation is not accounted for in Ref. \cite{Boulanger-Spindel2006}), and by cancelling the magnetic field ${\bf B}$ and assuming $a=1,\ b=0$ in the latter work---so that Eq. (52) of Ref. \cite{Boulanger-Spindel2006} corresponds to the Rindler metric
\be\label{Rindler}
ds^2=\left(1+\frac{g\zeta}{c^2} \right)^2 c^2dt^2-dx^2-dy^2-d\zeta ^2,
\ee 
relevant to a uniformly accelerated system. We find then an exact agreement if we substitute ${\bf g}\equiv (0,0,g)$ for ${\bf a}$ here, accounting for the fact that ${\bf S}=\hbar\Mat{\sigma }/2$, with $\hbar=1$ in Ref. \cite{Boulanger-Spindel2006}. This was expected, for the metric (\ref{Rindler}) coincides at PN order with the metric used by Hehl \& Ni \cite{HehlNi1990}, when $\Mat{\omega}={\bf 0}$. To get a description of the rotation effects using the approach of Ref. \cite{Boulanger-Spindel2006}, one could substitute the Kerr metric for the Rindler metric. We may also compare (\ref{2-spinor-HFW}) with the positive-energy PN Hamiltonian that appears implicitly in Eq. (67) of Ref. \cite{A38}, which corresponds to the metric
\be
ds^2=\left(1-\frac{U_\mathrm{N}({\bf r})}{c^2} \right)^2 c^2dt^2-\left(1+\frac{U_\mathrm{N}({\bf r})}{c^2} \right)^2(dx^2+dy^2+dz^2).
\ee\\
In Ref. \cite{A38}, the positive-energy states are automatically selected by the Ansatz (\ref{E-mc2}), as already noted. Naturally, the non-relativistic part in (\ref{2-spinor-HFW}) coincides with that in Ref. \cite{A38}, if we put $\Mat{\omega}={\bf 0}$ and substitute $U_\mathrm{N}={\bf g}.{\bf r}$ for ${\bf a}.{\bf r}$, with  ${\bf g}\equiv -{\bf a}$. (Cf. Note 1.) With this substitution, the two last (relativistic) terms in (\ref{2-spinor-HFW}) are identical to corresponding terms in Eq. (67) of Ref. \cite{A38}, up to positive factors, but the term with $\mathbf{a.r}\Delta$ is missing, and a relativistic term involving $\Delta U_\mathrm{N}$ has to be added. Of course, in that case, an exact agreement was not expected.\\

Let us now apply this to UCN on the rotating Earth. In Eq. (\ref{2-spinor-HFW}), ${\mathbf a}$ is then the absolute acceleration (denoted ${\bf a}_0$ in Note 1) of the non-inertial observer, which is at rest on the axis of the rotating frame. Thus \underline {${\bf a=0}$} in fact, as noted at the end of the foregoing Section. That is, if one computes the energy levels $E_\mathrm{nr}\equiv E-mc^2$ for the merely-rotating Earth, i.e., neglecting its gravitational field, the mere correction brought by using the PN Hamiltonian (\ref{2-spinor-HFW}) with ${\bf a=0}$ instead of the non-relativistic Hamiltonian (\ref{H-op}) with $V_0=0$, is the spin-rotation coupling term $-\Mat{\omega}.{\mathbf S}$. This is not really a relativistic effect, but one that arises due to the fact that the scalar Hamiltonian (\ref{H-op}) does not involve the intrinsic spin. In other words, the relativistic effects of the pure rotation are simply zero at the PN approximation. On the other hand, recall that the relativistic effects of the pure gravitational field of the Earth ({\it i.e.,} neglecting the rotation, but accounting for the {\it global} gravitational field) on the energy levels of UCN have been independently assessed \cite{A38,Boulanger-Spindel2006} and found fully negligible: they amount to less than $\simeq 3\times 10^{-53}$ J. Our net result is thus that all relativistic effects are negligible for UCN in the Earth's gravity.\\

The effects of the non-relativistic gravitational potential $-m{\bf g.x}$ are described in detail in Refs. \cite{Nesvizhevsky2003,WestphalAbele2006,VoroninAbele2006}, which, however, neglect the Earth's rotation. Hence, apart from the negligible relativistic effects, the perturbation to the stationary states as described in Refs. \cite{Nesvizhevsky2003,WestphalAbele2006,VoroninAbele2006} is due only to the terms $-\Mat{\omega}.{\mathbf L}$ and $-\Mat{\omega}.{\mathbf S}$ in the Hamiltonian (\ref{2-spinor-HFW}) which enters  the eigenstate equation (\ref{energy-state}). We have $\omega \simeq 7.27\times 10^{-5}\, \mathrm{s}^{-1}$. The spin-rotation coupling term $-\Mat{\omega}.{\mathbf S}$ (first noted by Mashhoon \cite{Mashhoon1988}) is hence approximately
\be
\mid \Mat{\omega}.{\mathbf S} \mid \simeq \omega \hbar \simeq 8\times 10^{-39}\,\mathrm{J},
\ee
which is still small, though less small, as compared with the non-relativistic energies $E_\mathrm{nr}$, that start from $1.4$ peV \cite{Nesvizhevsky2003}, which is {\it ca.} $2\times 10^{-31}$ J. However, it may be argued \cite{Mashhoon1995} that this coupling term has been indirectly measured in an experiment involving nuclear spin precession of atomic mercury \cite{Venema1992}.

\subsection{Effect of the angular-momentum term on the energy levels} \label{delta-energy}
Let us now estimate the effect of the angular-momentum term $\delta \mathrm{H}\equiv -\Mat{\omega}.{\mathbf L}$, which is the non-relativistic effect of the rotation, and is called ``Sagnac effect" \cite{WernerStaudenmannColella1979,KieferWeber2005,Mashhoon1988}. Taking cylindrical coordinates $\rho ,\theta ,\zeta $ with the axis $O\zeta $ being the rotation axis, thus parallel to vector $\Mat{\omega }$, we have
\be
-\Mat{\omega}.{\mathbf L}\varphi =i\hbar \Mat{\omega }.({\bf r} \wedge \nabla \varphi )
\ee
and
\be
\Mat{\omega}=\omega {\bf e}_\zeta, \quad {\bf r}=\rho {\bf e}_\rho +\zeta {\bf e}_\zeta , \quad \nabla \varphi = \frac{\partial\varphi}{\partial \rho } {\bf e}_\rho + \frac{1}{\rho }\frac{\partial\varphi}{\partial \theta  }  {\bf e}_\theta +\frac{\partial\varphi}{\partial \zeta  } {\bf e}_\zeta,
\ee
whence
\be \label{deltaH-phi}
(\delta \mathrm{H})\varphi\equiv -\Mat{\omega}.{\mathbf L}\varphi =i\hbar \omega \frac{\partial\varphi}{\partial \theta  }=i\hbar \omega \rho (\nabla \varphi ).{\bf e}_\theta.
\ee
To assess this, we take for $\varphi $ the unperturbed non-relativistic stationary state in the Earth's gravity \cite{Nesvizhevsky2003}, and we adopt Cartesian coordinates $x,y,z$ from point $A$, with radius vector ${\bf r}_0$, taken on the lower side of the slit [see after Eq. (\ref{H-op})], the axis $Ax$ being the West-East direction: ${\bf e}_x \equiv {\bf e}_\theta$, and the axis $Az$ being the local vertical. The stationary state is assumed to have the form \cite{WestphalAbele2006}
\be \label{separ-phi}
\varphi({\bf x}) = \varphi _v(z)e^{i(k_1x+k_2y)},
\ee
hence from (\ref{deltaH-phi}):
\be \label{deltaH-phi-stationary}
(\delta \mathrm{H})\varphi = i\hbar \omega \rho \frac{\partial\varphi}{\partial x  }=-\hbar \omega k_1 \rho \varphi = - \omega m u_1 \rho \varphi, 
\ee
with
\be
\qquad u_1\equiv {\bf u.e}_\theta, \quad {\bf u}\equiv \hbar(k_1{\bf e}_x+k_2{\bf e}_y)/m.
\ee
We note that the effect of the perturbing Hamiltonian $\delta \mathrm{H}\equiv -\Mat{\omega}.{\mathbf L}$ cannot be reduced to that of a mere additional potential $\delta V$, since, even for the particular wave functions (\ref{separ-phi}), the ratio $((\delta \mathrm{H})\varphi )/\varphi $ depends on $\varphi $ through $u_1\equiv \hbar k_1/m$. We note also that this relevant velocity $u_1$ is the {\it West-East velocity}. (Compare Ref. \cite{Granit0807}.)\\

In the neighborhood of the reference point $A$, the dependence of $\rho $ (the distance to the rotation axis) on the East-West coordinate $x$ is second-order. Calculating thus $\rho $ in the ``North-vertical" plane $Ayz$, we define the angle $\beta $ between $Az$ and the radius vector ${\bf AM}$. This adds to the latitude angle $\alpha $ to make the angle between the parallel to the equator and ${\bf AM}$ in the plane $Ayz$. We have $\cos \beta =z/(y^2+z^2)^{1/2}$, $\sin \beta =y/(y^2+z^2)^{1/2}$, hence we get ($R$ is the Earth's radius and $\rho _0\equiv R\cos \alpha$):
\be \label{rho-variable}
\rho = \rho _0 + (y^2+z^2)^{1/2}\cos(\alpha +\beta )=\rho _0+ z\cos\alpha -y\sin\alpha .
\ee
We compute the modification (\ref{deltaH-phi-stationary}) of the energy level as a first-order perturbation:
\be\label{delta-E}
(\delta E)_\mathrm{rot} \simeq (\varphi \mid (\delta \mathrm{H})\varphi)=-\omega m u_1 \int (\rho _0+ z\cos\alpha -y\sin\alpha)\abs{\varphi }^2\dd V,
\ee
with
\be \label{int}
\int (\rho _0+ z\cos\alpha -y\sin\alpha)\abs{\varphi }^2\dd V=\rho _0-\sin\alpha\int y \abs{\varphi}^2 \dd V +\cos\alpha \int z\abs{\varphi }^2\dd V.
\ee
Due to the form of the wave function, the first integral (with $y$) does not depend on the quantum state. Thus, in (\ref{int}), the contribution of the two first terms to $\delta E$ (\ref{delta-E}) is independent of the quantum state, hence may be omitted, because the energy levels are defined only up to a constant. That is, only the last term in (\ref{int}) contributes:
\be \label{delta-E-2}
(\delta E)_\mathrm{rot} \simeq -\omega m u_1 \cos\alpha \int z \abs{\varphi}^2 \dd V.
\ee

\paragraph{}
The integral in (\ref{delta-E-2}) is the average value $<z>_n$ of $z$ for the stationary state $\varphi =\varphi _n$. If we use reduced quantities,
\be \label{reduced}
\xi \equiv \frac{z}{l}, \quad \lambda \equiv \frac{E}{e}, \qquad l \equiv \left(\frac{\hbar^2}{2m^2 g}\right)^\frac{1}{3}, \quad e \equiv \left(\frac{\hbar^2 m g^2}{2}\right)^\frac{1}{3},
\ee
we have exactly \{\cite{Goodmanson2000}, eqs. (4b) and (18b)\}
\be
<\xi >_n = \frac{2}{3}\lambda _n.
\ee
By (\ref{reduced}), it follows  that
\be
<z>_n = \frac{2}{3}l\lambda _n=\frac{2}{3}\frac{E_n}{mg}=   \frac{2}{3}H_n,
\ee
where $H_n\equiv E_n/(mg)$ is the classical turning point \cite{VoroninAbele2006}. By (\ref{delta-E-2}), this implies that the {\it relative} modification of the energy level $E_n$ is independent of $E_n$:
\be \label{deltaEn}
\frac{(\delta E_n)_\mathrm{rot}}{E_n}\simeq -\frac{2\omega u_1 \cos\alpha}{3g}\simeq -3.4\times 10^{-5}
\ee
(for $u_1\simeq +10$ m/s and since $\omega \simeq 7.27\times 10^{-5}$ rad/s, $\cos\alpha\simeq 0.71$ and $g\simeq 10$ m/s$^2$). Hence, the effect of the Earth's rotation is not constant (as one finds by neglecting $\rho -\rho _0\simeq 10$ cm with respect to $\rho _0=R\cos \alpha \simeq 5\times 10^8$ cm, which seems justified at first sight), instead it is more important for high energy levels.
\footnote{
This was guessed by V. V. Nesvizhevsky (September 2006). His guess led me to do the calculation with a variable $\rho $, Eqs. (\ref{rho-variable})--(\ref{int}), giving the result (\ref{deltaEn}). Moreover, one referee of the present paper suggested that ``it might be helpful to provide an interpretation of the result," and the other one wrote that ``it is not a surprise that the modification of the energy level due to rotation increase with the energy, it just reflects the fact that centrifugal force increases with distance"---thus suggesting the explanation that follows.
}
The classical energy of the particle in the rotating frame of the Earth is \cite{L&LMeca}:
\be \label{E-conserved}
E_\mathrm{classical}=\frac{1}{2}m{\bf u}^2+V_0-\frac{1}{2}m (\Mat{\omega}\wedge {\bf x})^2=\frac{1}{2}m{\bf u}^2+mgz-\frac{1}{2}m \omega^2\rho ^2.
\ee
Thus, this result may be intuitively understood in saying that, when the altitude $z$ of the particle is increased, its distance $\rho $ to the rotation axis is increased proportionally: $\delta \rho =\delta z\,\cos\alpha \ll \rho $ from Eq. (\ref{rho-variable}).\\

Let us estimate the state number $n$ and the corresponding energy $E_n$ and height $H_n$, for which the modification $(\delta E_n)_\mathrm{rot}$ of the energy level due to the Earth's rotation becomes as large as the difference $\Delta E_n\equiv E_n-E_{n-1}$ between two successive energy levels. The WKB approximation, which is better at large $n$, gives $E_n \propto x_n^{2/3}$ with $x_n \equiv n-\frac{1}{4}$ \cite{WestphalAbele2006,VoroninAbele2006}, hence
\be
\Delta E_n \propto x_n^{-\frac{1}{3}} \left(1-\frac{1}{6x_n}\right)+O(x_n^{-\frac{7}{3}})\simeq Ax_n^{-\frac{1}{3}}, \qquad (\delta E_n)_\mathrm{rot} \simeq Bx_n^{\frac{2}{3}},
\ee
thus $\Delta E_n \simeq \abs{(\delta E_n)_\mathrm{rot}}$ when $x_n\simeq A/\abs{B}$. From $E_7\simeq 6.044$ and $E_6\simeq 5.431$ peV \cite{VoroninAbele2006}, we find $A\simeq 1.158$ peV and, using also (\ref{deltaEn}), $B\simeq -5.75\times 10^{-5}$ peV.  Hence
\be
\Delta E_n \simeq \abs{(\delta E_n)_\mathrm{rot}}\ \mathrm{when}\ n \simeq 20139 \simeq 20000,
\ee
corresponding to ({\it cf.} Table 1 in Ref. \cite{VoroninAbele2006})
\be
H_n \simeq H_7 \left(\frac{x_n}{x_7}\right)^{2/3}\simeq \left( 58.945 \times 207.25\right) \mathrm{\mu m}\simeq 12\, \mathrm{mm},
\ee
or
\be
E_n \simeq E_7 \left(\frac{x_n}{x_7}\right)^{2/3}\simeq \left(6.044 \times 207.25\right)\mathrm{peV} \simeq 1250\, \mathrm{peV}.
\ee
In the neutron trap, neutrons with different values and orientations of the horizontal velocity, thus with different values of the West-East velocity $u_1$ entering Eq.~(\ref{delta-E-2}), shall be present. Hence, for heights $h$ of the horizontal slit of the order of 12\,mm and larger, the higher-energy part of the corresponding spectrum, with $H_n \sim  h$, would be transformed to a continuum due to the Earth's rotation. 

\bigskip

{\bf Acknowledgement.} I did this work and [apart from redactional and editorial
 improvements and the interpretation based on Eq. (\ref{E-conserved})] I wrote this paper, during my participation to the Granit collaboration (i.e., between July 2006 and November 2006). Thus, for Subsect. \ref{delta-energy}, I benefitted from remarks made by the other participants, mainly remarks by V. V. Nesvizhevsky, K. V. Protasov, G. Pignol. Finally, the remarks of the referees allowed to improve the paper (in particular, see Note 2).

\end{document}